\documentclass[
 reprint,
 amsmath,
 amssymb,
 aps,
 preprintnumbers,
]{revtex4-1}

\usepackage{graphicx}
\usepackage{dcolumn}
\usepackage{bm}
\usepackage[caption=false]{subfig}
\usepackage{amsmath}
\usepackage{slashed}
\usepackage[utf8]{inputenc}

\DeclareMathOperator{\Tr}{Tr}
\newcommand{\Dslash}{\!\!\!\!/\,}
\newcommand{\half}{\frac{1}{2}}
\newcommand{\dt}{dt}
\newcommand{\ncfg}{100 }

\newcommand{\mpima}{0.4347(32) }
\newcommand{\mpimb}{0.5776(30) }
\newcommand{\mpimc}{0.6980(31) }

\newcommand{\mpika}{0.4349(43) }
\newcommand{\mpikb}{0.5769(40) }
\newcommand{\mpikc}{0.6987(40) }

\newcommand{\kapa}{0.13742726}
\newcommand{\kapb}{0.13703168}
\newcommand{\kapc}{0.13661366}

\newcommand{\mNma}{1.254(31) }
\newcommand{\mNmb}{1.371(21) } 
\newcommand{\mNmc}{1.500(18) }

\newcommand{\mNka}{1.195(31) }
\newcommand{\mNkb}{1.335(18) }
\newcommand{\mNkc}{1.479(16) }

\newcommand{\mRma}{2.236(79) }
\newcommand{\mRmb}{2.352(74) }
\newcommand{\mRmc}{2.446(68) }

\newcommand{\mRka}{2.366(76) }
\newcommand{\mRkb}{2.448(70) }
\newcommand{\mRkc}{2.532(65) }

\newcommand{\rchisqda}{0.619 }
\newcommand{\rchisqdb}{0.595 }
\newcommand{\rchisqdc}{0.842 }

\newcommand{\rchisqra}{1.002 }
\newcommand{\rchisqrb}{0.850 }
\newcommand{\rchisqrc}{0.757 }

\begin{document}

\preprint{ADP-19-22/T1102}

\title{Role of chiral symmetry in the nucleon excitation spectrum}

\author{Adam Virgili} \author{Waseem Kamleh} \author{Derek Leinweber}
\affiliation{Special Research Centre for the Subatomic Structure of
 	Matter, School of Physical Sciences, University of Adelaide,
 	South Australia 5005, Australia.}

\begin{abstract}
  
  The origin of the low-lying nature of the $N$*(1440), or Roper resonance, has been the subject of significant interest for many years, including several investigations using lattice QCD. The majority of lattice studies have not observed a low-lying excited state energy level in the region of the Roper resonance. However, it has been claimed that chiral symmetry could play an important role in our understanding of this resonance. The purpose of this study is to systematically examine the role of chiral symmetry in the low-lying nucleon spectrum by directly comparing the clover and overlap fermion actions. To ensure any differences in results are attributable to the choice of fermion action, simulations are performed on the same set of gauge field configurations at matched pion masses. Correlation matrix techniques are employed to determine the excitation energy of the first positive-parity excited state for each action. The clover and overlap actions show a remarkable level of agreement. We do not find any evidence that fermion action chiral symmetry plays a significant role in understanding the Roper resonance on the lattice.
  
\end{abstract}

\maketitle

\section{\label{sec:intro}Introduction}

The true nature of the Roper resonance ($N(1440){\frac{1}{2}}^+\, {P}_{11}$), the first positive-parity excited state of the nucleon discovered in 1964 via a partial-wave analysis of pion-nucleon scattering data~\cite{Roper:1964zza}, is a long standing source of debate. The puzzlement surrounding the Roper resonance arises from the discrepancy between the level ordering predicted by otherwise successful quark model calculations, and the energy of the resonance observed in nature. With an energy of 1440 MeV the Roper resonance is the lowest-lying resonance in the nucleon spectrum, sitting even below the first negative-parity excitation, the ($N(1535){\frac{1}{2}}^-\, {S}_{11}$) state. This is a reversal of the ordering predicted by simple quark models, which place the energy of the positive-parity $P_{11}$ state well above that of the negative-parity $S_{11}$ state.

This apparent discrepancy persists in lattice QCD calculations, with
the majority of lattice calculations obtaining an energy level for the
first positive-parity nucleon excitation that sits high relative to that expected for the Roper resonance, even near the physical quark mass regime~\cite{Mahbub:2010rm,Mahbub:2013ala,Liu:2016uzk,Kiratidis:2015vpa,Kiratidis:2016hda,Alexandrou:2014mka,Lang:2016hnn,Edwards:2011jj,Edwards:2012fx,Dudek:2012ag}. The exception to this is the $\chi$QCD Collaboration, which using overlap fermions
in combination with the sequential empirical Bayes (SEB) analysis
method~\cite{Chen:2004gp} were able to find a low-lying positive-parity
excited state on the lattice with an energy that in the chiral limit
is consistent with the Roper resonance in
nature~\cite{Liu:2014jua,Liu:2016rwa}.

It is clear that some controversy persists regarding how the Roper resonance in the continuum manifests on the lattice ~\cite{Eichmann:2016yit,Segovia:2016zyc,Eichmann:2016hgl,Liu:2016kbb,Yang:2017qan,Ren:2016xjm,Roberts:2016dnb,Segovia:2015hra}. The $\chi$QCD Collaboration advocates that their result is directly associated with the use of overlap fermions and stresses the importance of implementing exact chiral symmetry when investigating the nucleon spectrum. They motivate this by pointing towards the success of chiral soliton models, based on spontaneous chiral symmetry breaking, predicting the ordering of the Roper resonance and ${\rm{S}}_{11}$ state observed in nature and the contrasting failure of various, otherwise successful quark models, to do the same~\cite{Liu:2016rwa,Mathur:2003zf}.

Furthermore, motivated by the increased coupling of the overlap action to ghost states in the quenched approximation, it has been postulated the overlap action may provide better access to pion-nucleon physics on the lattice~\cite{Liu:2014jua}. On the other hand, even in the absence of a low-lying lattice energy level, it has been shown using effective field theory that it is possible to reconcile the lattice Wilson-type results with experiment by describing the infinite-volume Roper as a resonance generated dynamically through strongly coupled meson-baryon channels~\cite{Liu:2016uzk,Wu:2017qve}.

In light of these differing perspectives, it is important and of interest to perform a systematic investigation of the role of chiral symmetry in the nucleon spectrum~\cite{Padmanath:2019wid}. This is the aim of this study, where we directly compare results obtained from simulations respectively employing the overlap and clover fermion actions. To ensure that any discrepancies between the respective simulations are entirely attributable to the choice of fermion action, both simulations are performed on the same set of gauge field configurations, at matched pion masses, and analyzed utilizing identical correlation matrix techniques.

Calculations are performed at three values of the valence quark mass. The lattice energies of the ground and first positive-parity excited state are computed for each action from variational analyses, additionally yielding effective mass and eigenstate projected correlation functions which are also compared. Our final analysis avoids the selection of fit regimes, instead presenting the lattice results directly. 

The paper is structured as follows. Section~\ref{sec:methods} briefly reviews the background and methods used herein, and outlines the correlation matrix analysis technique. Section~\ref{sec:results} describes the simulation parameters and results, with conclusions presented in Sec.~\ref{sec:conclusions}.

\section{\label{sec:methods}Methodology}

\subsection{\label{subsec:resonancePhysics}Resonance Physics from Lattice QCD}

The determination of resonance properties from lattice QCD calculations requires a comprehensive
understanding of the spectrum of excited states in the finite periodic volume of the lattice.  In
principle, this spectrum includes all single, hybrid, and multiparticle contributions having the
quantum numbers of the resonance of interest.  This finite-volume spectrum then forms the input to
the L\"uscher method~\cite{Luscher:1990ux} or its
generalizations~\cite{Hall:2013qba,Liu:2016uzk,Wu:2017qve,Li:2019qvh} which relate the
finite-volume energy levels to infinite-volume momentum-dependent scattering amplitudes. The
application of these methods is a necessary step in connecting lattice QCD results to the
properties of resonances measured in experiment.
  
Obtaining an accurate determination of the finite-volume nucleon spectrum is challenging,
requiring an extensive set of baryon interpolating fields and robust correlation function analysis techniques.
Many collaborations have explored the spectrum excited by local single-particle
operators~\cite{Mahbub:2010rm,Mahbub:2013ala,Liu:2016uzk,Kiratidis:2015vpa,Kiratidis:2016hda, Alexandrou:2014mka,Lang:2016hnn,Edwards:2011jj,Edwards:2012fx}, and hybrid nucleon interpolators have been investigated in Ref.~\cite{Dudek:2012ag} where
additional states were found in the spectrum. Nonlocal multiparticle interpolating fields are necessary to determine the lattice energy eigenstates to the level of accuracy that is required to implement
the L\"uscher formalism and compare lattice QCD results to experiment.  Indeed, the main approach has
been to bring experimental results to the finite volume of the lattice \cite{Liu:2016uzk,Wu:2017qve}.
It is only recently that the first applications of the L\"uscher formalism to the baryon sector have
emerged~\cite{Andersen:2017una,Andersen:2019ktw}. 

In this work, our focus is on the finite-volume spectrum and its
dependence on the choice of fermion action as described in the
Introduction. We consider quasilocal operators as these are
sufficient to address this issue.  We acknowledge that these localized
operators do not have good overlap with multiparticle scattering
states and as such the energies obtained in our calculations may
contain contributions from more than one energy eigenstate of QCD.
The single-state ansatz of Ref.~\cite{Kiratidis:2015vpa} used herein
minimizes this effect, which can shift the observed energies within
the width of the associated resonance.  Moreover, we note that this
subtle issue of state mixing applies to both fermion actions. Any
differences which exist between the actions will be apparent in the
results presented, preserving our ability to search for a nontrivial
role for chiral symmetry in the nucleon spectrum.

\subsection{Lattice Fermion Actions}

Nucleon spectroscopy on the lattice is typically performed using Wilson fermions, with clover~\cite{Sheikholeslami:1985ij} and twisted mass~\cite{Frezzotti:2000nk} being the most common variants used today.
The (unimproved) Wilson fermion matrix is given by
\begin{equation}
  D_{\rm w} = \nabla\Dslash + \frac{a}{2}\Delta + m_{\rm w} \, ,
\end{equation}
where $\nabla$ is the central finite difference operator, $\Delta$ is the lattice Laplacian (or Wilson term), and $m_{\rm w}$ is the Wilson mass parameter~\cite{Wilson:1974sk}. 
The addition of the Wilson term successfully circumvents the infamous fermion doubling problem present in the naive lattice formulation. However it also explicitly violates chiral symmetry. The challenge in solving this dilemma is made manifest by the Nielson-Ninomiya no-go theorem~\cite{Nielsen:1980rz,Nielsen:1981hk,Nielsen:1981xu}, which prevents the existence of a doubler-free, local, chirally invariant formulation of fermions on the lattice that retains the correct continuum limit. The Ginsparg-Wilson relation~\cite{Ginsparg:1981bj}
\begin{equation}
  \{D,\gamma^5\} = 2aD\gamma^5D \, ,
\end{equation}
where $D$ is some lattice Dirac operator, offers a means to circumvent the no-go theorem by providing a softly broken implementation of chiral symmetry on the lattice.

Developed as a solution to the Ginsparg-Wilson relation, the overlap formalism~\cite{Narayanan:1992wx,Narayanan:1993sk,Narayanan:1993ss,Narayanan:1994gw,Neuberger:1997fp,Kikukawa:1997qh} is a formulation of fermions on the lattice which satisfies an exact, lattice-deformed chiral symmetry. The massless overlap Dirac operator is given by
\begin{equation}
D_o = \frac{1}{2a}(1+\gamma^5\epsilon(H_{\rm w})) \, ,
\end{equation}
where $\epsilon(H_{\rm w})$ is the matrix sign function applied to the kernel, $H_{\rm w}.$ Typically the kernel is chosen to be the Hermitian form of the Wilson Dirac operator, $H_{\rm w} = \gamma^5 D_{\rm w},$ but other choices are valid and in particular the use of a kernel that incorporates smearing can have numerical advantages~\cite{Kamleh:2001ff,Bietenholz:2002ks,Kovacs:2002nz,DeGrand:2004nq,Durr:2005mq,Durr:2005ik,Bietenholz:2006fj}. When used as a matrix kernel, the Wilson mass parameter $m_{\rm w}$ must be chosen to have a negative value in order to be in the topological region, with $am_{\rm w} = - 1$ being the canonical value at tree level~\cite{Hernandez:1998et,Neuberger:1999pz}.

Due to the presence of the matrix sign function, simulations which implement the overlap formalism are of the order of 100 times more computationally expensive than those with Wilson-type fermions. As such, it is far more common to employ Wilson fermions in a hadron spectrum calculation, where we expect the explicit breaking of chiral symmetry on the lattice to have a negligible impact.

However, one cannot immediately dismiss the possible role that the fermion action might play in examining the nucleon excitation spectrum. It is known that the coupling of interpolation fields to lattice hadron states is action dependent, and so it may be the case that a low-lying Roper-like state couples strongly with the overlap action but weakly with Wilson-type actions. 

The nonperturbatively improved clover action has been used extensively in previous studies of the nucleon excitation spectrum, and for this reason we choose to use this form of Wilson fermions in our comparison. The clover fermion matrix is given by
\begin{equation}
  D_{\rm cl} = \nabla\Dslash + \frac{a}{2}(\Delta  - \half c_{\rm sw} \sigma \cdot F) + m \, ,
\end{equation}
where $\sigma \cdot F$ is the clover term and $c_{\rm sw}$ is the clover coefficient, which can be nonperturbatively tuned to remove $O(a)$ errors. The quark mass for Wilson fermions is usually specified by the hopping parameter,
\begin{equation}
\kappa \equiv \frac{1}{8+2am} \, .
\end{equation}

In this work the overlap matrix kernel used is the fat-link irrelevant-clover (FLIC) fermion action~\cite{Kamleh:2001ff,Zanotti:2001yb},
\begin{equation}
D_{\rm flic} = \nabla\Dslash + \frac{a}{2}(\Delta^{\rm fl} - \frac{1}{2}\sigma\cdot F^{\rm fl}) + m_{\rm w} \, ,
\end{equation}
where the Wilson and clover terms are constructed from stout-smeared links with four sweeps of smearing at $\rho = 0.1.$

The massive overlap Dirac operator is defined as~\cite{Neuberger:1997bg}
\begin{equation}
D_o(\mu) = (1-\mu)D_o + \mu \, ,
\end{equation}
where $0 \leq \mu \leq 1$ is the overlap fermion mass parameter, representing a mass of $\frac{\mu}{1-\mu}.$

The external fermion propagator calculated using overlap fermions requires the subtraction of a contact term. After solving the linear system for a given fermion source $\psi,$
\begin{equation}
D_o(\mu)\chi = \psi \, ,
\end{equation}
each solution vector is modified as
\begin{equation}
\chi_{c} \equiv \frac{1}{2m_{\rm w}(1-\mu)}(\chi - \psi) \, ,
\end{equation}
in order to construct the external overlap quark propagator~\cite{Narayanan:1994gw,Edwards:1998wx},
\begin{equation}
S_{c} \equiv \frac{1}{2m_{\rm w}(1-\mu)}(D^{-1}_o(\mu) - 1) \, .
\end{equation}
Defining the bare mass $m^0$ via
\begin{equation}
m^0 = 2m_{\rm w}\mu \, ,
\end{equation}
through the above subtraction of the contact term, it is possible to show that
\begin{equation}
S^{-1}_{c} = S^{-1}_{c} \vert_{m^0 = 0} + m^0 \, . 
\end{equation}
and that exact chiral symmetry is obeyed
\begin{equation}
\{\gamma^5 ,S_{c}\vert_{m^0 = 0} \} = 0 \, ,
\end{equation}
just as in the continuum~\cite{Neuberger:1997fp}.

\subsection{\label{sec:cmt}Correlation matrix analysis}

Previous studies by the $\chi$QCD Collaboration~\cite{Liu:2014jua,Liu:2016rwa} obtained the nucleon excitation spectrum from overlap fermions with the SEB method~\cite{Chen:2004gp}. The majority of results from other groups use a variational analysis. Here we use the same correlation matrix method to extract the nucleon excitation spectrum for both actions in order to eliminate any other potential dependencies and perform a direct comparison of the results obtained from the clover and overlap fermion actions.

Variational correlation matrix techniques~\cite{Michael:1985ne,Luscher:1990ck} are well-established methods for successfully producing hadron spectra from correlation functions~\cite{Leinweber:2015kyz}. First, a basis of $N$ operators is chosen such that any states of interest are contained within the span. An $N \times N$ matrix of cross correlation functions,
\begin{equation}
\label{defn:CM}
\mathcal{G}_{ij}(\vec{p},t) = \sum_{\vec{x}}\textrm{e}^{-i\vec{p}\cdot\vec{x}}\,\big\langle\,\Omega\, \big| \,\chi_{i}(\vec{x},t)\,\overline{\chi}_{j}(\vec{0},t_{src})\, \big| \,\Omega\, \big\rangle \, ,
\end{equation}
is constructed, where $\overline{\chi}_{j}$ and $\chi_{i}$ are the creation and annihilation operators of the interpolating fields, respectively. The parity projection operator
\begin{equation}\label{ParityProjector}
\Gamma_{\pm} = \frac{1}{2}\,(\gamma_{0} \pm I)\, ,
\end{equation}
projects out definite parity at $\vec{p}=\vec{0}$. Defining $G_{ij}(\vec{p},\, t) = \Tr\left(\Gamma \, \mathcal{G}_{ij}(\vec{p},\, t)\right)$,  we
can write the Dirac-traced correlation function as a sum of exponentials,
\begin{equation}
\label{GijSum}
{G}_{ij}(t) = \sum_{\alpha}\lambda^{\alpha}_{i}\,\bar{\lambda}^{\alpha}_{j}\,\textrm{e}^{-m_{\alpha}t} \, ,
\end{equation}
where $\lambda^{\alpha}_{i}$ and $\bar{\lambda}^{\alpha}_{j}$ are the couplings of 
$\chi_{i}$ and $\overline{\chi}_{j}$ at the sink and source respectively, and $m_{\alpha}$ is the mass of the $\alpha \mathrm{th}$ energy eigenstate. We search for a
linear combination of operators
\begin{equation}
\bar{\phi}^{\alpha} = \bar{\chi}_{j}\,u^{\alpha}_{j} \qquad \textrm{ and } \qquad \phi^{\alpha} = \chi_{i}\,v^{\alpha}_{i} \, ,
\end{equation}
    such that $\phi$ and $\bar{\phi}$ ideally couple to a single energy
        eigenstate. In practice, the energies observed in lattice QCD calculations can be
        contaminated with states not captured by the spanned basis.  To minimize this effect,
        improved analysis techniques have been developed~\cite{Kiratidis:2015vpa}. From Eq.~(\ref{GijSum}) we see that
\begin{equation}
{G}_{ij}(t_{0} + \dt)\,u^{\alpha}_{j} = \textrm{e}^{-m_{\alpha}\dt}\,{G}_{ij}(t_{0})\,u^{\alpha}_{j} \, ,
\end{equation}
and note that we can now find $u^{\alpha}_{j}$ and $v^{\alpha}_{i}$ for a given choice of variational parameters $(t_0,\dt)$ by solving 
\begin{align}
\label{E-value-eq-L}
\big[{G}^{-1}(t_{0})\,{G}(t_{0} + \dt)\big]_{ij}\,u^{\alpha}_{j} &= c^{\alpha}\,u^{\alpha}_{i} \, ,
\end{align}
and 
\begin{align}
\label{E-value-eq-R}
v^{\alpha}_{i}\,\big[{G}(t_{0} + \dt)\,{G}^{-1}(t_{0})\big]_{ij} &= c^{\alpha}\,v^{\alpha}_{j} \, ,
\end{align}
the left- and right-handed eigenvalue equations with eigenvalue $c^{\alpha} = \textrm{e}^{-m_{\alpha}\dt}$. ${G}_{ij}$ is symmetric in the ensemble average so the improved estimator $\frac{1}{2}\,({G}_{ij} + {G}_{ji})$ is employed to ensure the left-handed and right-handed eigenvalues match. As ${G}_{ij}$ is diagonalized by $u^{\alpha}_{j}$and
$v^{\alpha}_{i}$ at $t_{0}$ and $t_{0} +\dt$ it is possible to write the eigenstate-projected correlation function as 
\begin{equation}
{G}^{\alpha}(t) = v^{\alpha}_{i}\,{G}_{ij}(t)\,u^{\alpha}_{j} \, .
\end{equation}
To extract eigenstate masses, we construct the effective mass function
\begin{equation}
  M_{\rm{eff}}^{\alpha}(t) = \ln \left( \frac{G^\alpha(t)}{G^\alpha(t+1)} \right) \, ,
  \label{eq:effmassfunc}
\end{equation}
and apply standard analysis techniques outlined in Ref.~\cite{Mahbub:2009nr}.

\section{\label{sec:results}Results}

\subsection{\label{subsec:simparams}Simulation Parameters}

Computations are performed on the $32^3 \times 64$ PACS-CS 2 + 1-flavor ensembles~\cite{Aoki:2008sm} at $\kappa = 0.13754$ providing a lattice spacing of $a = 0.0961$ fm and a sea quark mass corresponding to $m_\pi^2 = 0.1506(9)$ $\mathrm{GeV^2}$ in the Sommer scale with $r_0 = 0.49$ fm. The clover and overlap calculations employ identical sets of \ncfg configurations. Antiperiodic boundary conditions in time are applied for both actions. An operator basis is constructed for each action using 16, 35, 100, and 200 sweeps of Gaussian smearing~\cite{Gusken:1989qx} at the source and sink with smearing parameter $\alpha = 0.7$.

We select input parameters which minimize the computational cost of evaluating the matrix sign function of the overlap kernel $\epsilon(H)$, where $H\equiv\gamma_5D_{\rm flic}$ and $D_{\rm flic}$ is the FLIC fermion matrix~\cite{Kamleh:2001ff}. The Wilson term and clover links benefit from four sweeps of stout-link smearing at $\rho = 0.1.$ The Wilson mass parameter is set to $am_{\rm w} = -1.1,$ corresponding to a hopping parameter value of $\kappa = 0.17241$ in the kernel. The evaluation of the inner conjugate gradient is accelerated by projecting out the subspace corresponding to the $80$ lowest-lying eigenmodes of the overlap kernel and evaluating the sign function explicitly.  Overlap propagators are calculated at three values of the overlap mass parameter $\mu = 0.0628, \ 0.1205,$ and $0.1815,$ corresponding to pion masses of $m_\pi = \mpima, \ \mpimb,$ and $\mpimc$ GeV respectively. We note the lightest mass is similar to that considered in Ref.~\cite{Liu:2014jua}.

We compute the pion correlation functions for each action with 100 sweeps of source and sink smearing. To ensure the pion masses of the  respective actions match, we tune the clover hopping parameter by performing a linear fit to the square of the pion mass as a function of $1/\kappa$. Solving for the $\kappa$ values corresponding to the overlap pion masses, we obtain $\kappa = \kapa, \ \kapb,$ and $\kapc.$ The clover coefficient takes its nonperturbative value of $c_{\rm sw} = 1.715.$ Running the clover simulation with these tuned input parameters, we obtain pion masses which closely match those of the overlap simulation. Masses for both actions are presented in Table~\ref{t:pimass2}.

While the pion masses are carefully matched, both lattice fermion actions have $\mathcal{O}(a^2)$ errors that will lead to small discrepancies in the nucleon mass spectrum. However, these differences are small relative to the 300 MeV differences discussed in Ref.~\cite{Liu:2014jua}.

\begin{ruledtabular}

\begin{table}
  \centering
  \caption{Matched pion masses for the clover and overlap actions.}
  \begin{tabular}{  l c c c }
    \noalign{\smallskip}
		\multicolumn{2}{c}{Overlap} & \multicolumn{2}{c}{Clover} \\
                \noalign{\smallskip}
                \cline{1-2} \cline{3-4} 
                \noalign{\smallskip}
                $\mu$ & $m_\pi$/GeV & $\kappa$ & $m_\pi$/GeV \\
                \noalign{\smallskip}
                \hline
                \noalign{\smallskip}
                0.0628 & \mpima & \kapa & \mpika  \\
		0.1205 & \mpimb & \kapb & \mpikb  \\
		0.1815 & \mpimc & \kapc & \mpikc  \\                
	\end{tabular}
	\label{t:pimass2}
\end{table}

\end{ruledtabular}

\subsection{\label{subsec:cmanalysis}Correlation matrix analyses}

\begin{figure}[]
  \centering
  	\includegraphics[width=\linewidth]{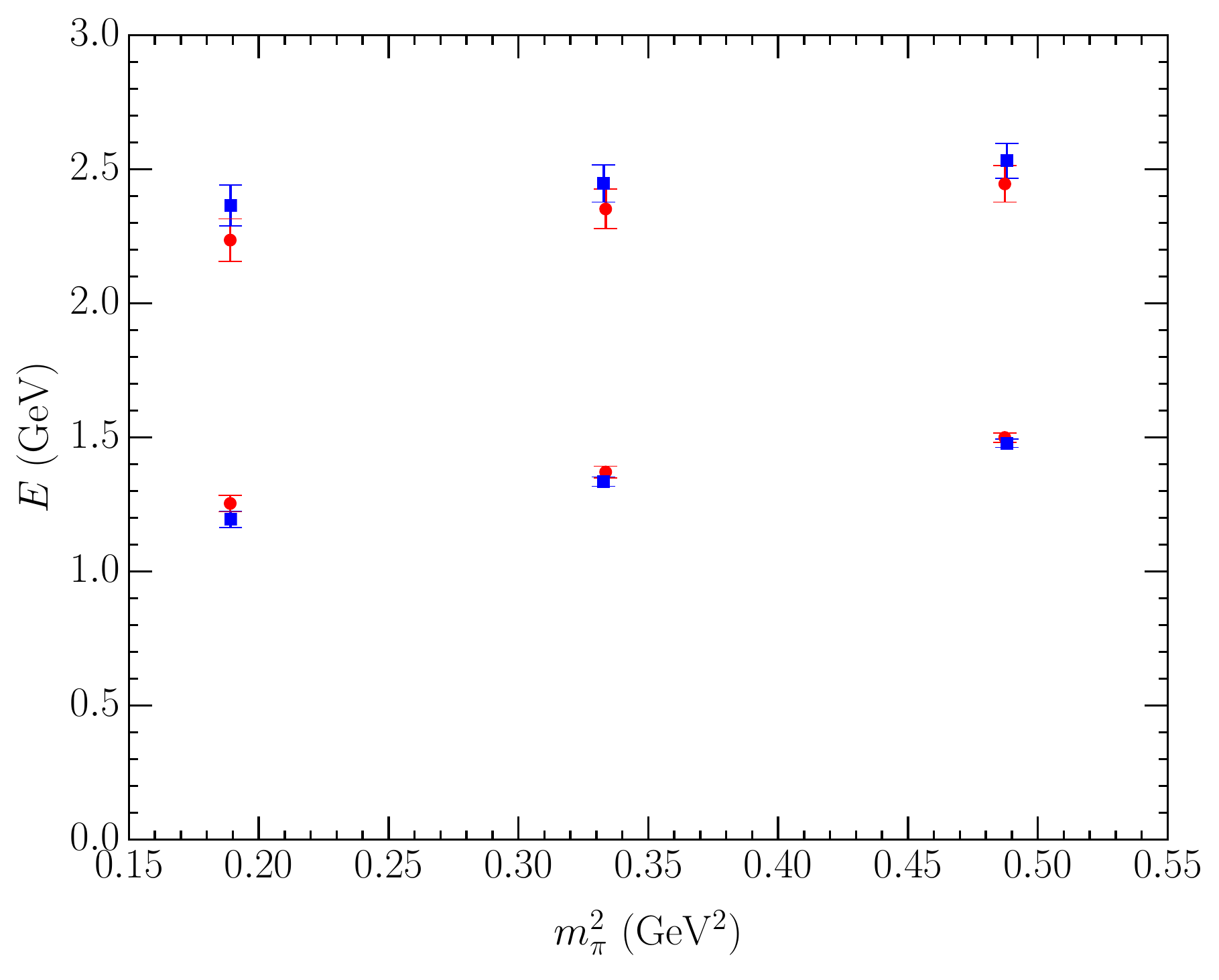}
	\caption{Nucleon ground and first-excited state masses for
          clover (blue, square) and overlap (red, circle) actions as a function of $m_\pi^2$.}
        \label{fig:masscompare}
\end{figure}

For our comparison, we employ the correlation matrix techniques discussed in Sec.~\ref{sec:cmt}. As we are only concerned with the ground state and first positive-parity excited state, we construct a $4\times4$ correlation matrix from our operator basis of source/sink Gaussian smearings ($N_{sm}$ = 16, 35, 100, 200) and select $t_0 = 1$ relative to the source at $t_s = 0$ and $dt=3$. Standard analysis techniques~\cite{Mahbub:2009nr} provide the results reported in Table~\ref{t:mass} and plotted in Fig~\ref{fig:masscompare}. All corresponding clover and overlap nucleon ground and first excited state masses are in statistical agreement. The small systematic differences are likely associated with the aforementioned $\mathcal{O}(a^2)$ errors in the fermion action.

These results are dependent on specific choices for the variational parameters and fit windows. To make our results more robust we investigate further, initially avoiding the selection of fit windows. Here we compare the eigenstate projected effective mass and correlation functions for each action. Specifically, we compare how the first excited state compares to the ground state for each action without fits. We do this in two ways.
 
First, we consider the effective mass functions obtained from the eigenstate projected correlators as in Eq.(\ref{eq:effmassfunc}). We then take the ratio
\begin{equation}
  R_{\alpha/0}(t) = M_{\mathrm{eff}}^\alpha(t)/M_{\mathrm{eff}}^0(t) \, ,
\end{equation}
for each action, where $M_{\mathrm{eff}}^0$ and $M_{\mathrm{eff}}^\alpha$ are the ground and $\alpha \mathrm{th}$ excited state effective mass functions. This scales the excited state mass function for each action in terms of their respective ground states, eliminating any differences which arise from ground state discrepancies and placing the focus on the excitation energy. To compare the actions we take the ratio
\begin{equation}
  \mathcal{R}_\alpha(t) = \frac{R_{\alpha/0}^{\mathrm{clover}}(t)}{R_{\alpha/0}^{\mathrm{overlap}}(t)} \, .
\end{equation}
  
As a second point of comparison, we consider the eigenstate projected correlation functions directly. We take the ratio
\begin{equation}
  G_{\alpha/0}(t) = G^\alpha(t)/G^0(t) \, ,
\end{equation}
for each action, where $G^0$ and $G^\alpha$ are the ground and $\alpha \mathrm{th}$ excited state projected correlation functions. We construct the effective mass splitting
\begin{equation}
  \Delta M^\alpha_{\mathrm{eff}}(t) = \ln \left( \frac{G_{\alpha/0}(t)}{G_{\alpha/0}(t+1)} \right) \, ,
\end{equation}
corresponding to the mass splitting of the $\alpha \mathrm{th}$ excited state and the ground state for each action, respectively. Taking the difference
\begin{equation}
  \mathcal{D}_\alpha(t) =  \Delta M_{\mathrm{eff}}^{\alpha,\mathrm{clover}}(t) - \Delta M_{\mathrm{eff}}^{\alpha,\mathrm{overlap}}(t) \, ,
\end{equation}
we obtain the difference between the mass splittings produced by each action.

\begin{figure}[t!]
	\centering
	\includegraphics[width=\linewidth]{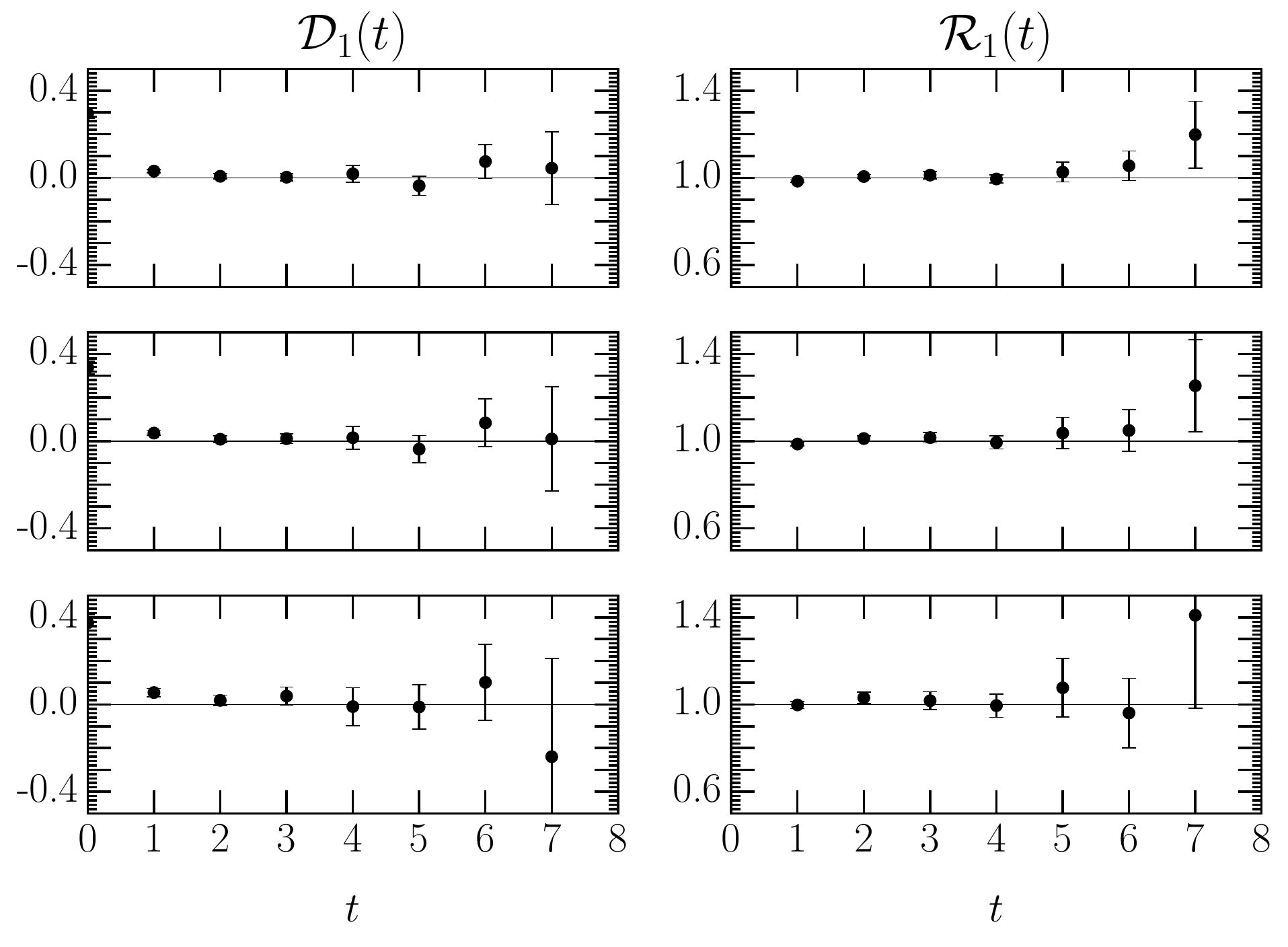}

	\caption{$\mathcal{D}_1(t)$ in units of GeV (left) and $\mathcal{R}_1(t)$ (right) for three different valence quark masses with heaviest $\mu = 0.1815$, $\kappa = \kapc$ (top), $\mu = 0.1205$, $\kappa = \kapb$ (middle), and lightest $\mu = 0.0628$, $\kappa = \kapa$ (bottom), for variational parameters $t_0 = 1$, $t= t_0 + dt = 4$.}
	\label{fig:MDR}
\end{figure}

Both $\mathcal{D}_1(t)$ and $\mathcal{R}_1(t)$ are plotted in Fig.~\ref{fig:MDR} for each mass regime. We note that $\mathcal{D}_\alpha(t) = 0$ or $\mathcal{R}_\alpha(t) = 1$ correspond to no difference in the excitation energies produced by the clover and overlap fermion actions.

It is important to demonstrate that these highly correlated ratios and differences can be described with the full covariance matrix $\chi^2/\mathrm{d.o.f.} \approx 1$. Hence, we evaluate the agreement of the difference $\mathcal{D}_1(t)$ and the ratio $\mathcal{R}_1(t)$ with the constants zero and one, respectively. The reduced $\chi^2$ values of each fit with $2 \leqslant t \leqslant 6$ are reported in Table~\ref{t:MDR}. These values confirm there is negligible difference between the clover and overlap actions with respect to the nucleon spectrum, and show no evidence for the existence of a low-lying lattice excited state.

\begin{figure}[t!]
	\centering
	\includegraphics[width=\linewidth]{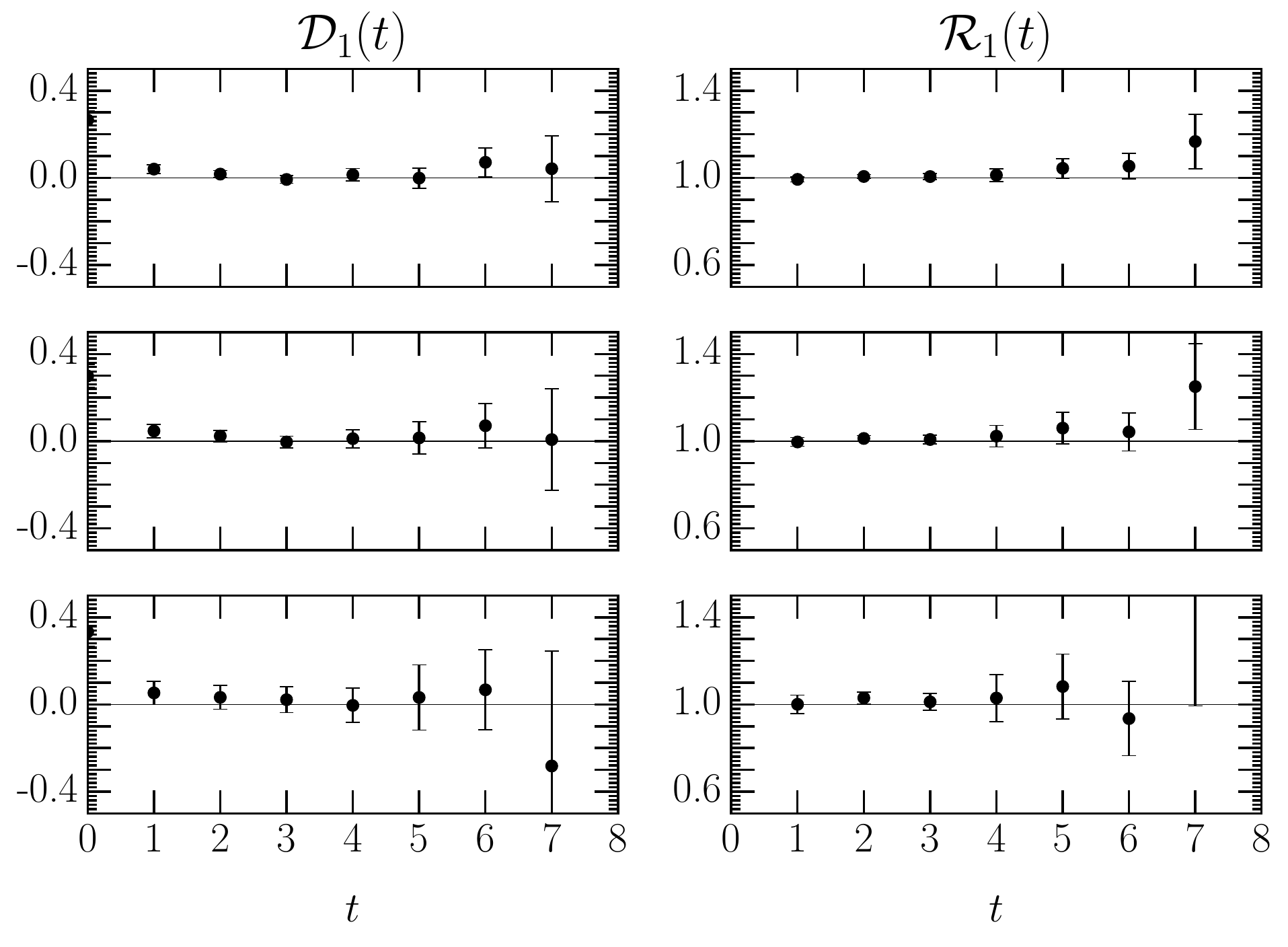}

	\caption{$\mathcal{D}_1(t)$ in units of GeV (left) and $\mathcal{R}_1(t)$ (right) for three different valence quark masses with heaviest $\mu = 0.1815$, $\kappa = \kapc$ (top), $\mu = 0.1205$, $\kappa = \kapb$ (middle), and lightest $\mu = 0.0628$, $\kappa = \kapa$ (bottom), for variational parameters $t_0 = 2$, $t = t_0 + dt = 5$.}
	\label{fig:MDR2}
\end{figure}

The next step in our comprehensive analysis is to explore other variational parameters $t_0$ and $dt$. Systematic errors in the correlation matrix analysis enter as $\mathcal{O}(e^{-(E_{N+1} - E_{N})t})$ for an $N \times N$ correlation matrix. To this end we require a large $t = t_0 + dt$. However, we also require a small $t_0$ to ensure statistically accurate information is captured from excited state contributions before they are Euclidean-time suppressed. Table~\ref{t:MDR:varparams} presents the $\chi^2/\mathrm{d.o.f.}$ values for the results discussed in Table~\ref{t:MDR}, this time focusing on the lightest quark mass, closest to the sea quark mass. Again, analyses of $\mathcal{D}_1(t) = 0$ and $\mathcal{R}_1(t) = 1$ return acceptable $\chi^2/\mathrm{d.o.f.}$ values. Figure~\ref{fig:MDR2} displays results corresponding to Fig.~\ref{fig:MDR}, this time for the variational parameters $t_0=2$, $dt=3$.

\begin{figure}[t!]
	\centering
	\includegraphics[width=\linewidth]{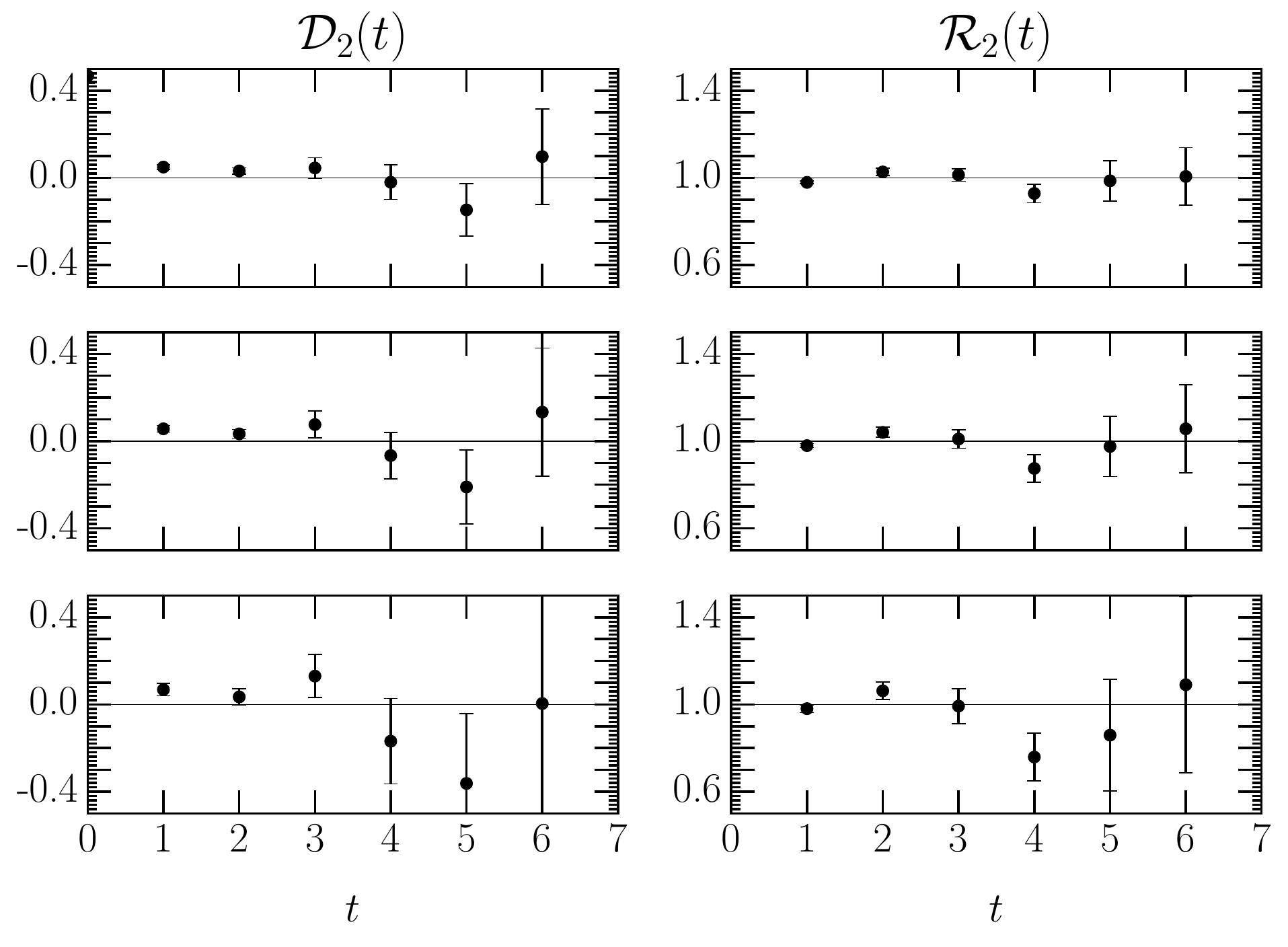}

	\caption{$\mathcal{D}_2(t)$ in units of GeV (left) and $\mathcal{R}_2(t)$ (right) for three different valence quark masses with heaviest $\mu = 0.1815$, $\kappa = \kapc$ (top), $\mu = 0.1205$, $\kappa = \kapb$ (middle), and lightest $\mu = 0.0628$, $\kappa = \kapa$ (bottom), for variational parameters $t_0 = 1$, $t = t_0 + dt = 4$.}
	\label{fig:MDR3}
\end{figure}

Finally, we consider the second excited state.  The
    energy of the second excited state is $\sim$3
    GeV~\cite{Mahbub:2010rm}. With the limited number of
    configurations used here, the correlation functions decay to
    noise rapidly with our $a=0.0961$ fm lattice spacing.
    Nonetheless, a solution
    of the generalized eigenvalue problem is found for variational
    parameters $t_0$ = 1, $dt$ = 3 and results for $\mathcal{D}_2(t)$
    and $\mathcal{R}_2(t)$ are presented in Fig.~\ref{fig:MDR3}.
    These results are similar to those presented for
    $\mathcal{D}_1(t)$ and $\mathcal{R}_1(t)$.  While the statistical
    fluctuations are notably larger, there is no evidence of a
    significant difference between the fermion actions.

\begin{ruledtabular}
\begin{table}
  \centering
  \caption{Masses in GeV of the ground state and the first positive-parity excited state for the clover and overlap actions at the three valence quark masses considered.}
  \begin{tabular}{  l c c }
    \noalign{\smallskip}
    & Overlap & Clover \\
    \noalign{\smallskip}
    \hline
    \noalign{\smallskip}
    Ground State & \mNma & \mNka \\
    & \mNmb & \mNkb \\
    & \mNmc & \mNkc \\
    First Excited State & \mRma & \mRka \\
    & \mRmb & \mRkb \\
    & \mRmc & \mRkc \\
  \end{tabular} 
  \label{t:mass}
\end{table}
\end{ruledtabular}

\begin{ruledtabular}
\begin{table}[]
	\centering
        \caption{$\chi^2/\mathrm{d.o.f.}$ for $\mathcal{D}_1(t)$ fitted to the constant 0 and $\mathcal{R}_1(t)$ fitted to the constant 1, for each quark mass regime with $2 \leqslant t \leqslant 6$.}
        \begin{tabular}{l c c c}
		\noalign{\smallskip}
		\multicolumn{2}{c}{} & \multicolumn{2}{c}{$\chi^2/\mathrm{d.o.f.}$} \\
                \noalign{\smallskip}
                \cline{3-4}
		\noalign{\smallskip}
                $\mu$ & $\kappa$ & $\mathcal{D}_1(t)$ & $\mathcal{R}_1(t)$ \\
                \noalign{\smallskip}
                \hline
                \noalign{\smallskip}
		0.0628 & \kapa & \rchisqda & \rchisqra  \\
		0.1205 & \kapb & \rchisqdb & \rchisqrb  \\
		0.1815 & \kapc & \rchisqdc & \rchisqrc  \\

	\end{tabular}
	\label{t:MDR}
\end{table}
\end{ruledtabular}

\begin{ruledtabular}
\begin{table}
  \centering
   \caption{$\chi^2/\mathrm{d.o.f.}$ for $\mathcal{D}_1(t) = 0$ and $\mathcal{R}_1(t) = 1$, for the lightest quark mass considered with variational parameters $t_0$ and $t=t_0+dt$ relative to the source at $t_s=0$.}
   \label{t:MDR:varparams}
	\begin{tabular}{c c c c}
		\noalign{\smallskip}
		\multicolumn{2}{c}{} & \multicolumn{2}{c}{$\chi^2/\mathrm{d.o.f.}$} \\
		\noalign{\smallskip}
                \cline{3-4}
                \noalign{\smallskip}
                $t_0$ & $t$ & $\mathcal{D}_1(t)$ & $\mathcal{R}_1(t)$ \\
                \noalign{\smallskip}
                \hline
                \noalign{\smallskip}
		1 & 4 & \rchisqda & \rchisqra  \\
		1 & 5 & 0.493 & 0.930  \\
                2 & 4 & 0.816 & 0.962  \\
		2 & 5 & 0.528 & 0.843  \\
	\end{tabular}
\end{table}
\end{ruledtabular}

\section{\label{sec:conclusions}Conclusions}

In this paper, the role of chiral symmetry in the nucleon excitation spectrum was systematically examined. Results obtained from simulations employing nonchiral clover fermions and chiral overlap fermions were compared. To ensure that any observed differences or discrepancies in the results are attributable to the choice of action the simulations were performed on the same set of gauge field configurations at three matched pion masses.

All corresponding clover and overlap nucleon ground and first excited state masses are in statistical agreement. Further analysis was conducted, showing that the ratios of the first excited and ground state effective mass functions and mass splittings are the same for each action.

The results show a remarkable level of agreement between the clover and overlap actions. Hence, we do not find any evidence supporting the claim that chiral symmetry plays a significant role in understanding the Roper resonance on the lattice.

\begin{acknowledgments}

This research was supported by resources and services provided by the
National Computational Infrastructure (NCI) in Canberra, Australia,
which is supported by the Australian Government; The Pawsey
Supercomputing Centre in Perth, Australia, which is funded by the
Australian Government and the Government of Western Australia; and the
Phoenix HPC service at the University of Adelaide. This research was
supported by the Australian Research Council through ARC Discovery
Project Grants Nos.\ DP150103164, DP190102215, and DP190100297 and ARC
Linkage Infrastructure Grant LE190100021.

\end{acknowledgments}

\bibliographystyle{apsrev4-1}
\bibliography{Bibliography}

\end{document}